\documentclass[preprintnumbers,amsmath,amssymb,nofootinbib,longbibliograpphy]{revtex4-2}
\usepackage{dcolumn}
\usepackage{mathrsfs}
\usepackage{amsmath}
\usepackage{graphicx}
\usepackage{bm}
\usepackage{epstopdf}
\usepackage{float}
\usepackage{hyperref}
\usepackage{color}
\usepackage{amssymb}
\usepackage{eucal}
\usepackage{mathrsfs}
\usepackage{amsthm}
\usepackage{changes}

\def\bk{{\bf k}}
\def\bkp{{\bf k'}}
\def\uk{{\hat{\bk}}}
\def\br{{\bf r}}
\def\brp{{\bf r'}}
\def\wk{\omega_k}
\def\wkp{\omega_{k'}}
\def\wA{\omega_A}
\def\w0{\omega_0}
\def\ak{a_{\bk}}
\def\bA{{\bf A}}
\def\bE{{\bf E}}
\def\bB{{\bf B}}
\def\bR{{\bf R}}
\def\bRp{{\bf R'}}

\def\bmu{{\bm \mu}}

\def\akd{a_{\bk}^\dagger}
\def\ekr{e^{i\bk \cdot \br}}
\def\emkr{e^{-i\bk \cdot \br}}
\def\ewt{e^{i\wk t}}
\def\emwt{e^{-i\wk t}}
\def\akl{a_{\bk \lambda}}
\def\akld{a_{\bk \lambda}^\dagger}
\def\ekl{\hat{{\bf e}}_{\bk \lambda}}
\def\eklp{\hat{{\bf e}}_{\bkp \lambda '}}
\def\eklnb{\hat{e}_{\bk \lambda}}
\def\eklpnb{\hat{e}_{\bkp \lambda '}}

\newcommand{\sgn}{\operatorname{sgn}}

\begin{document}

\title{Nonlocal Static and Dynamical Vacuum Field Correlations and Casimir-Polder Interactions}

\author{Roberto Passante$^{1,2,}$\footnote{roberto.passante@unipa.it}}
\author{Lucia Rizzuto$^{1,2,}$\footnote{lucia.rizzuto@unipa.it}}

\affiliation{$^1$Dipartimento di Fisica e Chimica - Emilio Segr\`{e},
Universit\`{a} degli Studi di Palermo, Via Archirafi 36, I-90123 Palermo,
Italy}
\affiliation{$^2$Istituto Nazionale di Fisica Nucleare, Laboratori Nazionali del Sud, I-95123 Catania, Italy}

\begin{abstract}
In this review we investigate several aspects and features of spatial field correlations for the massless scalar field and the electromagnetic field, both in stationary and nonstationary conditions, and show how they manifest in two- and many-body static and dynamic dispersion interactions (van der Waals and Casimir-Polder). We initially analyze the spatial field correlations for noninteracting fields, stressing their nonlocal behavior, and their relation to two-body dispersion interactions. We then consider how field correlations are modified by the presence of a field source, such as an atom or in general a polarizable body, firstly in a stationary condition and then in a dynamical condition, starting from a nonstationary state. We first evaluate the spatial field correlation for the electric field in the stationary case, in the presence of a ground-state or excited-state atom, and then we consider its time evolution in the case of an initially nonstationary state. We discuss in detail their nonlocal features, in both stationary and nonstationary conditions. We then explicitly show how the nonlocality of field correlations can manifest itself in van der Waals and Casimir-Polder interactions between atoms, both in static and dynamic situations. We discuss how this can allow to indirectly probe the existence and the properties of nonlocal vacuum field correlations of the electromagnetic field, a research subject of strong actual interest, also in consequence of recent measurements of spatial field correlations exploiting electro-optical sampling techniques. The subtle and intriguing relation between nonlocality and causality is also discussed.
\end{abstract}

\maketitle

\section{\label{sec:1} Introduction}

In quantum field theory, field commutators vanish outside the light cone, as expected from relativistic causality considerations \cite{Schwartz14}. However, it is well known that vacuum field correlations in general do not vanish for space-like separations; this is not in contradiction with relativistic causality since a nonvanishing field correlation function does not allow possibility of a superluminal transmission of information \cite{Kaku93,Biswas-Compagno90,Passante-LaBarbera95,Petrosky-Ordonez00,Franson08}.

The fact that quantum vacuum fluctuations, for example in the case of the electromagnetic field, possess nonvanishing correlations between spacetime points separated by a space-like interval can however have profound conceptual implications and physical consequences. Examples are inducing correlations between quantum objects in causally disconnected regions of spacetime \cite{Settembrini-Lindel22,Biswas-Compagno90} and determining radiation-mediated (van der Waals and Casimir-Polder) interactions between atoms \cite{Power-Thirunamachandran93a,Power-Thirunamachandran93b,Passante-Persico03,Salam10,Passante18}, even in dynamical (time-dependent) conditions \cite{Passante-Persico06}.
Dispersion interactions are thus an indirect evidence of the existence of vacuum fluctuation and of their spectral and spatial correlation features. Direct measurements of the vacuum fluctuations of the quantum electric field have been reported, as well as their dependence on the space-time volume sampled \cite{Riek-Seletskiy15}.

Recently, some experiments have been able to demonstrate the existence of vacuum field correlations between non-causally connected space-time points, exploiting electro-optic sampling \cite{Settembrini-Lindel22,BeneaChelmus-Settembrini19}. Previous theoretical achievements on vacuum fluctuations and their spatial correlations, and the experimental possibility of proving their existence, make very important to analyze in detail their properties and physical consequences, as well as their relevance on observable physical effects. Very recently, the possibility of experimentally probe and individually separate vacuum field fluctuations and source radiation contributions exploiting electro-optic sampling has been also proposed \cite{Lindel-Harter23}.

The experimental results and theoretical proposals above are currently giving a strong theoretical and experimental boost and relevance to studying the physical features of vacuum field correlations, including their nonlocality, in both stationary and nonstationary conditions, as well as their effects on different quantum systems and observable quantities.

The existence of zero-point field fluctuations is one of the most prominent prediction of the quantum theory of the electromagnetic field, and quantum fields in general; even in the ground state of the free field, the electric and magnetic fields fluctuate around their vanishing average value, as a consequence of the Heisenberg uncertainty principle resulting from the noncommutativity of components of the electric and magnetic field operators \cite{Compagno-Passante05}. These fluctuations are at the origin of many important observable effects, such as the spontaneous decay of an excited atom in vacuum, the Lamb shift, and the Casimir and Casimir-Polder forces between neutral objects in vacuum \cite{Milonni94,Compagno-Passante05,Craig-Thirunamachandran98,Salam10,Milonni19}.

In this paper we review relevant properties of the electromagnetic and scalar field vacuum fluctuations, both in the vacuum and in the presence of field sources. We consider both equilibrium and nonequilibrium conditions. Specifically, we investigate and review their space-time correlations in both static and dynamical conditions, their role in static and dynamical radiation-mediated effects such as two- and many-body dispersion interactions (van der Waals and Casimir-Polder), as well as their nonlocal properties and consequent relation with entanglement between field and matter observables \cite{Power-Thirunamachandran93a,Power-Thirunamachandran93b,Cirone-Passante96,Cirone-Passante97,Passante-Persico05,Passante-Persico06,Passante-Persico07,Salam09}.

This paper is organized as follows. In section \ref{sec:2} we first introduce the equal-time vacuum field correlations in the field ground state, both for the massless scalar field and the electromagnetic field, stressing their nonlocal features. We then discuss how they are modified by the presence of a polarizable body such as a ground- or excited-state atom, considering a stationary configuration. In section \ref{sec:3} we investigate how Casimir-Polder dispersion interactions between two or three atoms give an indirect evidence of the nonlocal features of the spatial field correlations. In section \ref{sec:4} we consider nonstationary cases, when the interacting system starts from a nonequilibrium configuration, discussing in detail the time evolution of the nonlocal spatial correlations of the electric field. We then investigate how this leads to dynamical three-body Casimir-Polder interactions, both in the case of three bare ground-state atoms and when one of the three atoms is initially in a bare excited state. Section \ref{sec:5} is devoted to our conclusive remarks.

\section{\label{sec:2} Stationary vacuum field correlations}

We first discuss the vacuum field correlations for the massless relativistic scalar field and for the electromagnetic field, in a stationary case, specifically the vacuum state.
Let us first consider a massless real scalar field; the field operator is given by \cite{Mandl-Shaw10}

\begin{equation}\label{eq:scalar field}
\phi (\br ,t) = \sum_\bk \left( \frac {\hbar c^2}{2V\wk} \right)^{1/2} \left(  \ak \ekr \emwt + \akd \emkr \ewt \right) ,
\end{equation}
where $\wk = c \lvert \bk \rvert$, $V$ is the quantization volume, the annihilation and creation operators $\ak$ and $\akd$ satisfy the usual bosonic commutation rules, and periodic boundary conditions are used.
We have $\langle 0 \lvert \phi (\br ,t) \rvert 0 \rangle$, where $\rvert 0 \rangle$ is the vacuum state of the field defined as $\ak \rvert 0 \rangle =0$ for all $\bk$. The equal-time vacuum correlation function is easily obtained as

\begin{equation}\label{scalar correlation function discrete}
\langle 0 \lvert \phi (\br ,t) \phi (\brp ,t) \rvert 0 \rangle = \frac {\hbar c^2}{2V} \sum_\bk \frac 1{\wk} e^{i\bk \cdot (\br -\brp )} .
\end{equation}

In the continuum limit, $V \rightarrow \infty$, $\sum_\bk \rightarrow V/(2\pi )^3\int d^3k$, Eq. (\ref{scalar correlation function discrete}) becomes

\begin{equation}\label{corrsf1}
\langle 0 \lvert \phi (\br ,t) \phi (\brp ,t) \rvert 0 \rangle = \frac {\hbar c}{4\pi} \frac {1}{2i} \frac 1{\lvert \br - \brp \rvert} \lim_{\eta \to 0^+} \int_0^\infty
\left( e^{ik(\lvert \br - \brp \rvert +i\eta )} - e^{-ik(\lvert \br - \brp \rvert -i\eta )} \right) ,
\end{equation}
where an exponential regularization factor $e^{-\eta k}$, with $\eta >0$, has been introduced. After the integration over $k$, taking $\eta \rightarrow 0$, we have (for $\br \neq \brp$)

\begin{equation}\label{corrsf2}
\langle 0 \lvert \phi (\br ,t) \phi (\brp ,t) \rvert 0 \rangle = \frac {\hbar c}{4\pi} \frac 1{\lvert \br - \brp \rvert^2} .
\end{equation}

This relation shows the existence of nonlocal spatial field correlations outside the light cone, as well as a divergence on the light cone (even if Eq. (\ref{corrsf2}) is an equal-time correlation function, it can be easily extended to the general case).

Similar considerations hold for the electromagnetic field, that is the main point of this paper. We are mainly interested to the nonrelativistic quantum electrodynamical case, and in this case it is convenient to work in the Coulomb gauge, $\nabla \cdot \bA = 0$. The expressions of the vector potential, electric field and magnetic field operators are (Gauss units) \cite{Compagno-Passante05}

\begin{eqnarray}\label{emfieldoperators}
\bA (\br ,t) &=&  \sum_{\bk \lambda} \bA (\bk \lambda ; \br ,t) = \sum_{\bk \lambda} \left( \frac {2\pi \hbar c^2}{\wk V} \right)^{1/2} \ekl \left( \akl (t) \ekr + \akld (t) \emkr \right) ,
\nonumber \\
\bE (\br ,t) &=& \sum_{\bk \lambda} \bE (\bk \lambda ; \br ,t) = i\sum_{\bk \lambda} \left( \frac {2\pi \hbar \wk}{V} \right)^{1/2} \ekl \left( \akl (t) \ekr - \akld (t) \emkr \right) ,
\nonumber \\
\bB (\br ,t) &=& \sum_{\bk \lambda} \bB (\bk \lambda ; \br ,t) = i\sum_{\bk \lambda} \left( \frac {2\pi \hbar \wk}{V} \right)^{1/2} (\uk \times \ekl ) \left( \akl (t) \ekr - \akld (t) \emkr \right) ,
\end{eqnarray}
where $\hat{\bk}=\bk /\lvert \bk \rvert$ is a unit vector along the direction of the wavevector $\bk$, $\ekl$ ($\lambda =1,2$) are polarization unit vectors, assumed real, and $\wk = ck$.

In the vacuum state, defined by $\akl \lvert 0 \rangle =0$ for any $(\bk \lambda )$, we have $\langle \bE (\br ,t) \rangle = 0$, $\langle \bB (\br ,t) \rangle = 0$, and the vacuum equal-time correlation function for the Fourier components of the (transverse) electric field is easily found as

\begin{equation}\label{em EE  F correlation function}
\sum_\lambda \langle 0 \lvert E_i (\bk \lambda ; \br ,t) E_j (\bk \lambda ; \brp ,t) \rvert 0 \rangle = \frac {2\pi \hbar \wk}{V}\sum_{\lambda} (\eklnb )_i (\eklnb )_j e^{i\bk \cdot (\br -\brp )} =
\frac {2\pi \hbar \wk}{V} \left(\delta_{ij} - \hat{k}_i \hat{k}_j \right) e^{i\bk \cdot \bR },
\end{equation}
where $\bR =\br -\brp$ and from now on the Einstein's convention of repeated indices is used. For the sum over the polarizations, we have used

\begin{equation}\label{polarization sum}
\sum_{\lambda =1,2} (\eklnb )_i (\eklnb )_j = \delta_{ij} - \hat{k}_i \hat{k}_j .
\end{equation}

In the continuum limit, $V \rightarrow \infty$, $\sum_\bk \rightarrow V/(2\pi )^3 \int \! d^3k$, after integration over $\bk$, we obtain the equal-time vacuum spatial correlation for the components of the electric field

\begin{equation}\label{em EE correlation function}
\langle 0 \lvert E_i (\br ,t) E_j (\brp ,t) \rvert 0 \rangle = \frac {2\pi \hbar \wk}{V}\sum_{\bk} \left(\delta_{ij} - \hat{k}_i \hat{k}_j \right) e^{i\bk \cdot (\br -\brp )} = - \frac {4\hbar c}{\pi} \left( \delta_{ij} - 2 \hat{R}_i \hat{R}_j \right) \frac 1{R^4} ,
\end{equation}
where $\hat{R}_i$ is a component of the unit vector $\hat{\bR} = \bR /\lvert \bR \rvert$.

Similarly to the massless scalar field case, Eq. (\ref{em EE correlation function}) shows that vacuum fluctuations of the electric field have nonlocal spatial correlations (i.e., they do not vanish outside the light cone) \cite{Biswas-Compagno90,Spagnolo-Dalvit07}.

There are also vacuum nonlocal correlations between components of the magnetic field, as well as between components of the electric and magnetic fields. For the magnetic-magnetic correlation, we have

\begin{eqnarray}\label{em MM  F correlation function}
\sum_\lambda \langle 0 \lvert B_i (\bk \lambda ; \br ,t) B_j (\bk \lambda ; \brp ,t) \rvert 0 \rangle &=& \frac {2\pi \hbar \wk}{V}\sum_{\lambda} (\hat{\bk} \times \ekl )_i (\hat{\bk} \times \ekl )_j e^{i\bk \cdot (\br -\brp )}
\nonumber \\
&\ & =\frac {2\pi \hbar \wk}{V} \left(\delta_{ij} - \hat{k}_i \hat{k}_j \right) e^{i\bk \cdot \bR },
\end{eqnarray}
where, for the sum over the polarizations, we have used

\begin{equation}\label{polarization sum 2}
\sum_{\lambda =1,2} (\hat{\bk} \times \ekl )_i (\hat{\bk} \times \ekl )_j = \delta_{ij} - \hat{k}_i \hat{k}_j .
\end{equation}
Eq. (\ref{em MM  F correlation function}), after summation over $\bk$ in the continuum limit gives, for the spatial correlation of the magnetic field, the same expression (\ref{em EE correlation function}) for the electric field case

\begin{equation}\label{em BB correlation function}
\langle 0 \lvert B_i (\br ,t) B_j (\brp ,t) \rvert 0 \rangle = - \frac {4\hbar c}{\pi} \left( \delta_{ij} - 2 \hat{R}_i \hat{R}_j \right) \frac 1{R^4} .
\end{equation}

Finally, for the electric-magnetic case we get

\begin{equation}\label{em EB  F correlation function}
\sum_\lambda \langle 0 \lvert E_i (\bk \lambda ; \br ,t) B_j (\bk \lambda ; \brp ,t) \rvert 0 \rangle =
\frac {2\pi \hbar \wk}{V}\sum_{\lambda} (\ekl )_i (\hat{\bk} \times\ekl )_j e^{i\bk \cdot (\br -\brp )} =
\frac {2\pi \hbar \wk}{V} \epsilon_{ij\ell} \hat{k}_\ell e^{i\bk \cdot \bR },
\end{equation}
where for the sum over polarizations we have used

\begin{equation}\label{polarization sum 3}
\sum_{\lambda =1,2} (\ekl )_i (\hat{\bk} \times \ekl )_j = \epsilon_{ij\ell} \hat{k}_\ell ,
\end{equation}
and $\epsilon_{ij\ell}$ is the totally antisymmetric Levi-Civita symbol \cite{Craig-Thirunamachandran98}. Analogously,

\begin{equation}\label{em BE  F correlation function}
\sum_\lambda \langle 0 \lvert B_j (\bk \lambda ; \brp ,t) E_i (\bk \lambda ; \br ,t) \rvert 0 \rangle =
\frac {2\pi \hbar \wk}{V} \epsilon_{ij\ell} \hat{k}_\ell e^{-i\bk \cdot \bR } .
\end{equation}

In the continuum limit, the angular $\bk$ integration of (\ref{em EB  F correlation function}) yields

\begin{equation}
\label{em EB  F correlation function 2}
\int d\Omega_k \sum_\lambda \langle 0 \lvert E_i (\bk \lambda ; \brp ,t) B_j (\bk \lambda ; \br ,t) \rvert 0 \rangle =
-\frac {8\pi^2 i \hbar \wk}{kV} \epsilon_{ij\ell} \nabla_\ell \left( \frac {\sin (kr)}{kr} \right) ,
\end{equation}
that is purely imaginary \cite{Spagnolo-Dalvit07}. The electric-magnetic correlation function (\ref{em BE  F correlation function}), after polarization sum and angular integration, yields the same expression of the first one, Eq. (\ref{em EB  F correlation function 2}), but with the opposite sign. Therefore the symmetrized e-m vacuum correlation function vanishes.

In the next section we will discuss the deep relation between vacuum field correlation functions, also in the presence of field sources or in dynamical situations, with van der Waals and Casimir-Polder interactions between atoms or in general electrically and magnetically polarizable bodies.

Spatial field correlations are modified by the presence of boundary conditions \cite{Power-Thirunamachandran82,Spagnolo-Passante06} or field sources, an atom or a polarizable body for example \cite{Cirone-Passante96,Cirone-Passante97}. In the latter case we speak of {\it dressed} spatial field correlations. Let us assume that an atom A is placed at $\br_A$, and $\bmu_A$ is its dipole moment operator. The Hamiltonian of the system, in the multipolar coupling scheme and within the dipole approximation, is

\begin{equation}\label{Hamiltonian one atom}
H = H_A + H_F - \bmu_A \cdot \bE (\br_A) ,
\end{equation}
where $H_A$ and $H_F$ are respectively the free atom and the free field Hamiltonians, $\bmu_A$ is the dipole moment operator of the atom A and $\bE (\br_A)$ is the electric field operator evaluated at the position $\br_A$ of the atom (it is indeed the transverse displacement field, that outside the field source coincides with the total electric field) \cite{Compagno-Passante05,Salam10,Andrews-Bradshaw20,Stokes-Ahsan22,Buhmann-I13}. We just mention that also macroscopic boundary condition yield changes to the spatial correlations of the electromagnetic field and related physical phenomena \cite{Power-Thirunamachandran82,Buhmann-I13,Buhmann-II13,Spagnolo-Passante06,Ford22}. Very recently, spatial field correlations of the massless scalar field between points at the opposite side of a movable perfect mirror (thus subjected to quantum fluctuations of its position) have been investigated \cite{Montalbano-Armata23}.

Due to the atom-field interaction, the non-interacting ground state $\lvert g_A, 0 \rangle$, where $\lvert g_A \rangle$ indicates the ground state of the atom A and $\lvert 0 \rangle$ the vacuum field state, is not an eigenstate of the total Hamiltonian $H$. At the second order in the atom-field coupling, the true (interacting) ground state $\lvert \tilde{g}_A\rangle$ of the system can be obtained by time-independent perturbation theory in the form \cite{Power-Thirunamachandran93a,Cirone-Passante97,Passante18}

\begin{eqnarray} \label{dressed ground state}
\lvert \tilde{g}_A \rangle &=& (1+N) \lvert g_A, 0 \rangle -i\left( \frac {2\pi}{\hbar V}\right)^{1/2} \sum_n \sum_{\bk \lambda}
\frac {\wk^{1/2}(\bmu^{ng}_A \cdot \ekl )e^{-i\bk \cdot \br_A}}{\omega_{ng}+\wk} \lvert n, 1_{\bk \lambda} \rangle
\nonumber \\
&\ & -\frac {2\pi}{\hbar V} \sum_{mn} \sum_{\bk \bkp \lambda \lambda '}
\frac {(\wk \wkp )^{1/2}(\bmu^{nm}_A \cdot \eklp ) (\bmu^{mg}_A \cdot \ekl ) e^{-i(\bk +\bkp )\cdot \br_A} }{(\wk +\omega_{mg})(\wk +\wkp +\omega_{ng})}
\lvert n, 1_{\bk \lambda} 1_{\bkp \lambda '} \rangle ,
\end{eqnarray}
where
$\lvert n, 1_{\bk \lambda} \rangle$ and $\lvert n, 1_{\bk \lambda} 1_{\bkp \lambda '} \rangle$ are eigenstates of $H_0$ in the form of a tensor product of the atomic state $n$ and one- and two-photon Fock states, respectively; also,

\begin{equation}\label{normfactor}
N = -\frac{\pi \hbar}V \sum_n \sum_{\bk \lambda} \frac {\wk \lvert \bmu^{ng}_A \cdot \ekl \rvert^2}{(\omega_{ng}+\wk )^2}
\end{equation}
is a normalization factor. In the above equations, $m$, $n$ indicate a complete set of atomic states, $\bmu^{mn}_A = \langle m \lvert \bmu_A \rvert n \rangle$ are the matrix elements of the electric dipole moment operator of atom A, and $\omega_{mg} = (E_m-E_g)/\hbar$ are atomic transition frequencies from the ground state. Equation (\ref{dressed ground state}) shows that the dressed ground state is not separable in the atom and field space state, and the atom is surrounded by a cloud of virtual photons \cite{Passante-Power87,Maclay22}. A similar situation occurs for a physical nucleon where, in the framework of generalized parton distributions, the bare nucleon is surrounded by a cloud of virtual mesons, and an expression analogous to (\ref{dressed ground state}) is indeed obtained \cite{Pasquini-Boffi06}.

We can now evaluate the equal-time correlation function between modes of the electric field \cite{Cirone-Passante97}

\begin{eqnarray}\label{dressed static correlation ground}
&\ & \langle \tilde{g}_A \rvert E_i(\bk \lambda ;\br ,t) E_j(\bkp \lambda '; \brp ,t)\lvert \tilde{g}_A \rangle = \frac {2\pi \hbar \wk}{V} (\eklnb )_i (\eklnb )_j e^{i\bk \cdot (\br -\brp )} \delta_{\bk \bkp} \delta_{\lambda \lambda '}
\nonumber \\
&\ & + \frac {4\pi^2}{V^2} \wk \wkp \sum_n  (\mu_A^{gn})_p (\mu_A^{ng})_q (\eklnb )_i (\eklpnb )_j (\eklnb )_q  (\eklpnb )_p
\left[ \frac {e^{-i\bk \cdot (\br -\br_A)}{e^{i\bkp \cdot (\brp -\br_A)}}}{(\wk + \omega_{ng})(\wkp + \omega_{ng})} \right.
\nonumber \\
&\ & \left. +\frac {e^{i\bk \cdot (\br -\br_A)}{e^{i\bkp \cdot (\brp -\br_A)}}}{\wk +\wkp} \left( \frac 1{\wk +\omega_{ng}}+  \frac 1{\wkp +\omega_{ng}} \right) + \text{c.c.}
\right] ,
\end{eqnarray}
where c.c. stands for the complex conjugate and, for the sake of simplicity and, without loss of generality, we have assumed that the dipole matrix elements are real, that is $\bmu^{ng}=\bmu^{gn}$.

The first term in the RHS of (\ref{dressed static correlation ground}), apart from the polarization sum, is essentially the same in (\ref{em EE  F correlation function}), that is the {\it bare} correlation function, while the second term gives the contribution due to the presence of atom A, yielding the {\it dressed} correlation function. Eq. (\ref{dressed static correlation ground}) shows that in the presence of an atom, contrarily to the bare vacuum case, also different field modes become correlated. In the next section we will discuss in detail the importance of this point in the non-additive three-body van der Waals and Casimir-Polder dispersion interaction between ground state atoms \cite{Cirone-Passante97,Cirone-Passante96,Passante-Persico05}.

Summation of (\ref{dressed static correlation ground}) over the field modes yields the dressed spatial correlation function of the complete electric field at two different points of space and equal time

\begin{equation}
\label{correlation electric field ground 1}
\begin{split}
\langle \tilde{g}_A \rvert E_i(\br ) E_j(\brp ) \lvert  \tilde{g}_A \rangle & = \sum_{\bk \bkp \lambda \lambda '} \langle \tilde{g}_A \rvert E_i(\bk \lambda ;\br ,t) E_j(\bkp \lambda '; \brp ,t)\lvert \tilde{g}_A \rangle
\\
& = \langle 0 \rvert E_i(\br ) E_j(\brp ) \lvert  0 \rangle + \langle \tilde{g}_A \rvert E_i(\br ) E_j(\brp ) \lvert  \tilde{g}_A \rangle_D ,
\end{split}
\end{equation}
where the first term is the same given in Equation (\ref{em EE correlation function}), that is the bare correlation function, and the second term is the (dressing) correction due to the presence of the ground-state atom A. The latter term involves polarization sum (see (\ref{polarization sum})) and angular integrations that can be easily performed using
\begin{equation}
\label{angular integration}
\int d\Omega \left( \delta_{pq} -\hat{k}_p \hat{k}_q \right) e^{\pm i\bk \cdot \bR} = \frac 1{k^2} F_{pq}^R \int d\Omega  e^{\pm i\bk \cdot \bR} = \frac {4\pi}{k^3}F_{pq}^R \frac {\sin (kR)}{R} ,
\end{equation}
where we have defined the following differential operator

\begin{equation}\label{differential operator}
F_{pq}^R = \left( -\delta_{pq} \nabla^2 +\nabla_p \nabla_q \right)^R ,
\end{equation}
and the apex $R$ indicates the coordinate with respect to which the derivatives are taken. After straightforward algebra, we obtain the second term in (\ref{correlation electric field ground 1}) at the second order in the atom-field coupling in the form of a double frequency integration

\begin{equation}\label{correlation electric field ground 2}
\begin{split}
& \langle \tilde{g}_A \rvert E_i(\br ) E_j(\brp ) \lvert  \tilde{g}_A \rangle_D
= \frac 1{\pi^2} \sum_n (\mu_A^{gn})_p (\mu_A^{ng})_q
F_{qi}^R F_{pj}^{R'} \frac 1{RR'} \int_0^\infty d\wk \int_0^\infty d\wkp
\\
& \times  \left[  \frac 1{\omega_{ng}+\wk} \left( \frac 1{k'-k} + \frac 1{k'+k} \right) + \text{c.c.} \right]
\left( \sin (kR) \sin (k'R') + \sin (k'R) \sin (kR') \right) ,
\end{split}
\end{equation}
where $\bR =\br -\br_A$ and $\bRp = \brp -\br_A$. Performing first the integration over $\wkp$ and then that over $\wk$, we finally obtain

\begin{equation}\label{correlation electric field ground 3}
\begin{split}
\langle \tilde{g}_A \rvert E_i(\br ) E_j(\brp ) \lvert  \tilde{g}_A \rangle_D
& = \frac 2{\pi^2} \sum_n (\mu_A^{gn})_p (\mu_A^{ng})_q F_{qi}^R F_{pj}^{R'} \frac 1{RR'} \int_0^\infty d\wk \frac {\sin [\omega_{ng} (R+R')/c]}{\wk +\omega_{ng}}
\\
& = \frac 2\pi \sum_n (\mu_A^{gn})_p (\mu_A^{ng})_q
F_{qi}^R F_{pj}^{R'} \frac 1{RR'} f[\omega_{ng} (R+R')/c] ,
\end{split}
\end{equation}
where $f(z)= \text{ci}(z) \sin (z) - \text{si}(z)\cos (z)$ is the auxiliary function of the sine and cosine integral functions, $\text{si}(z)$ and $\text{ci}(z)$ respectively \cite{NISTHandbook10}.
Eq. (\ref{correlation electric field ground 3}) shows that the dressing part of the correlation function is monotonically decreasing with the distance, with a characteristic distance scale given by $c/\omega_{ng}$. Asymptotically, for $R,R' \gg c/\omega_{ng}$, we have $f(\omega_{ng} (R+R')/c) \sim (\omega_{ng} (R+R')/c)^{-1}$ \cite{NISTHandbook10}. Thus, eq. (\ref{correlation electric field ground 3}) asymptotically scales with the inverse seventh power of the distance from the atom A; this $R^{-7}$ distance scaling should be compared with the $R^{-4}$ scaling of the bare vacuum fluctuations in (\ref{em EE correlation function}). In the next section we will discuss the relevance of this property for the non-additive three-body Casimir-Polder interaction between ground-state atoms or molecules in the far-zone (retarded) regime.

Equations (\ref{dressed ground state}-\ref{dressed static correlation ground}) have been obtained for a ground-state atom. If we consider the case of an atom in one of its excited states, in the continuum limit there is an extra contribution from the frequency integrations due to the resonance pole \cite{Passante-Persico05}. In this case, for simplicity we consider a two-level atom with transition frequency $\wA =(E_e-E_g)/\hbar$, located at $\br_A$, and interacting with the quantum electromagnetic field through the Hamiltonian (\ref{Hamiltonian one atom}). $\lvert g_A \rangle$ and $\lvert e_A \rangle$ indicate the ground and excited state of atom A, with energy $E_g$ and $E_e$ respectively. The bare excited state is then $\lvert e_A, 0 \rangle$. We assume to consider timescales shorter than the lifetime of the excited state, and thus we neglect the spontaneous decay of the atom. The second-order interacting excited state is then \cite{Power-Thirunamachandran93a,Passante-Persico05}

\begin{eqnarray} \label{dressed excited state}
\lvert \tilde{e}_A \rangle &=& (1+N') \lvert e_A, 0 \rangle +i\left( \frac {2\pi}{\hbar V}\right)^{1/2} \sum_{\bk \lambda}
\frac {\wk^{1/2}(\bmu^{ge}_A \cdot \ekl )e^{-i\bk \cdot \br_A}}{\wA-\wk} \lvert g_A, 1_{\bk \lambda} \rangle
\nonumber \\
&\ & +\frac {2\pi}{\hbar V} \sum_{\bk \bkp \lambda \lambda '}
\frac {(\wk \wkp )^{1/2}(\bmu^{ge}_A \cdot \eklp ) (\bmu^{ge}_A \cdot \ekl ) e^{-i(\bk +\bkp )\cdot \br_A} }{(\wA - \wk )(\wk +\wkp )}
\lvert e_A, 1_{\bk \lambda} 1_{\bkp \lambda '} \rangle ,
\end{eqnarray}
where $N'$ is a normalization factor. The presence of a pole at $\wk =\wA$, related to the possibility of a real transition from the excited to the ground state, should be noted; we will see that its presence has important consequences on the spatial correlations of the field and on the three-body Casimir-Polder interaction when an excited atom is involved.

Analogously to the ground-state case, we can now evaluate the spatial correlation function relative to two modes of the electric field on the dressed state (\ref{dressed excited state}), and we obtain, up to the second order in the atom-field coupling (we are assuming that the matrix elements $\bmu_A^{eg}$ are real) \cite{Passante-Persico05}

\begin{eqnarray}\label{dressed static correlation excited}
&\ & \langle \tilde{e}_A \rvert E_i(\bk \lambda ;\br ,t) E_j(\bkp \lambda '; \brp ,t)\lvert \tilde{e}_A \rangle = \frac {2\pi \hbar \wk}{V} (\eklnb )_i (\eklnb )_j e^{i\bk \cdot (\br -\brp )} \delta_{\bk \bkp} \delta_{\lambda \lambda '}
\nonumber \\
&\ & + \frac {4\pi^2}{V^2} \wk \wkp (\mu_A^{eg})_p (\mu_A^{eg})_q (\eklnb )_i (\eklpnb )_j (\eklnb )_q  (\eklpnb )_p
\left[ \frac {e^{i\bk \cdot (\br -\br_A)}{e^{-i\bkp \cdot (\brp -\br_A)}}}{(\wA -\wk )(\wA -\wkp )} \right.
\nonumber \\
&\ & \left. -\frac {e^{i\bk \cdot (\br -\br_A)}{e^{i\bkp \cdot (\brp -\br_A)}}}{\wk +\wkp} \left( \frac 1{\wA -\wk}+  \frac 1{\wA -\wkp} \right) + \text{c.c.}
\right] .
\end{eqnarray}
With a procedure analogous to the previous ground-state case we can now obtain, in the continuum limit, the following expression for the spatial correlation function of the complete electric field

\begin{equation}
\label{correlation electric field excited 1}
\begin{split}
\langle \tilde{e}_A \rvert E_i(\br ) E_j(\brp ) \lvert  \tilde{e}_A \rangle & = \sum_{\bk \bkp \lambda \lambda '} \langle \tilde{e}_A \rvert E_i(\bk \lambda ;\br ,t) E_j(\bkp \lambda '; \brp ,t)\lvert \tilde{e}_A \rangle
\nonumber \\
& = \langle 0 \rvert E_i(\br ) E_j(\brp ) \lvert  0 \rangle + \langle \tilde{e}_A \rvert E_i(\br ) E_j(\brp ) \lvert  \tilde{e}_A \rangle_D,
\end{split}
\end{equation}
where $ \langle 0 \rvert E_i(\br ) E_j(\brp ) \lvert  0 \rangle$ is the bare spatial correlation (\ref{em EE correlation function}) and the dressing correlation
$\langle \tilde{e}_A \rvert E_i(\br ) E_j(\brp ) \lvert  \tilde{e}_A \rangle_D$ has a structure analogous to that of the ground-state case previously considered. The main difference is the presence of a pole at $\wk ,\wkp = \wA$ in the frequency integration path, as it is evident from Eq. (\ref{dressed static correlation excited}), yielding a resonant contribution to the correlation function. The final result is \cite{Passante-Persico05}

\begin{equation}
\label{correlation electric field excited 2}
\begin{split}
& \langle \tilde{e}_A \rvert E_i(\br ) E_j(\brp ) \lvert  \tilde{e}_A \rangle_D  = \frac 2{\pi} (\mu_A^{eg})_q (\mu_A^{eg})_p
\\
&\times F_{qi}^R F_{pj}^{R'} \frac 1{RR'} \left[\text{PV} \int_0^\infty \! d\omega \frac {\sin [\omega (R+R')/c]}{\omega -\wA}
+2\pi \sin \left( \frac{\wA R}{c} \right) \sin \left( \frac{\wA R'}{c} \right) \right] ,
\end{split}
\end{equation}
where PV indicates the Principal Value of the integral. This equation contains two terms. Comparison of (\ref{correlation electric field excited 2}) with (\ref{correlation electric field ground 3}) shows that the first term has a similar structure as in the case of the ground state and includes contributions from all field modes; the second term in (\ref{correlation electric field excited 2}) is a new one, originating from the resonant pole at $\omega = \wA$ and shows spatial oscillations with a scale given by $c/\wA$ \cite{Passante-Persico05}.

\section{\label{sec:3} Bare and dressed vacuum field correlations and stationary Casimir-Polder dispersion interactions between atoms}

In this section we explore the deep connection between spatial correlations of the vacuum electromagnetic field and dispersion interactions (van der Waals and Casimir-Polder) between atoms, including many-body effects. The physical basis is that, even in the vacuum state, the electric and magnetic fields fluctuate around their zero average value, and they induce instantaneous dipole moments in electrically and/or magnetically polarizable objects such as atoms or molecules: since vacuum fields are spatially correlated, as we have shown in the previous section, the induced atomic dipole moments will be correlated too, and this eventually leads to an interaction energy between them \cite{Power-Thirunamachandran93b,Salam10,Passante18}. Furthermore, since these correlations change when a third atom is present (see previous section), its presence affects the interaction between the two other atoms, leading to non-additive effects in the dispersion interaction between three or more atoms \cite{Passante-Persico05,Salam10}. We will now address these effects.

We first analyze the two-body interaction case.

\subsection{Two-body dispersion interactions}
\label{Two-body dispersion interactions}
Let consider two ground-state atoms or in general two polarizable bodies B and C, located in $\br_B$ and $\br_C$, respectively. For simplicity, we assume they are isotropic. We now discuss an heuristic model to obtain the dispersion interactions between two atoms based on vacuum field correlations \cite{Power-Thirunamachandran93b,Salam10,Buhmann-I13}; we would like to mention that this heuristic model can be also rigorously derived from the standard atom-field Hamiltonian \cite{Passante-Persico03}. In the two points where atoms B and C are located, fluctuations of the electric and magnetic fields exists, and, as shown in the previous sections, these field fluctuations are spatially correlated; thus, for each field mode, they induce and correlate dipole moments in the two atoms. We have, in the spirit of a linear response theory \cite{Power-Thirunamachandran93b,Salam10,Passante18,Milonni19,Passante-Rizzuto21},

\begin{equation}\label{induced dipole}
\bmu_{B(C)i}^e(\bk ) \sim \alpha_{B(C)}^e(\wk ) E_i(\bk \lambda ; \br_{B(C)}), \, \, \, \,
\bmu_{B(C)i}^m(\bk ) \sim \alpha_{B(C)}^m(\wk ) B_i(\bk \lambda ; \br_{B(C)})
\end{equation}
where $\bmu_{B(C)i}^e(\bk )$ and $\bmu_{B(C)i}^m(\bk )$ are respectively the Fourier components of the induced electric and magnetic dipole moments of one of the two atoms; also,
$\alpha_{B(C)}^e(\wk )$ and $\alpha_{B(C)}^m(\wk )$ are, respectively, the electric and magnetic dynamical polarizabilities of the atoms, that we assume isotropic so that they are scalar functions of the frequency. We also assume that the induced and correlated dipoles interact via a classical dipole-dipole interaction between oscillating dipole moments. Thus, we have

\begin{equation}\label{interaction correlated dipoles}
\Delta E^{ab} = \sum_{\bk \lambda} \langle \mu_{Bi}^{a}(\bk ) \mu_{Cj}^{b}(\bk ) \rangle V_{ij}^{.ab}(k , \bR = \br_C -\br_B)
\end{equation}
where $a,b=e \, \text{(electric)},\, m \, \text{(magnetic)}$ denote the electric or magnetic components, and $V_{ij}^{ab}(k , \bR)$ is the classical potential tensor between electric or magnetic dipole moments
\cite{Spagnolo-Dalvit07,Salam10}. This tensor, representing the tensor part of the interaction energy between two dipoles oscillating at a frequency $\wk =ck$ and averaged over an oscillation period, for the electric-electric part is \cite{Power-Thirunamachandran93b,Spagnolo-Dalvit07,Salam10}

\begin{equation}\label{potential tensor ee}
\begin{split}
V_{ij}^{ee}(k , \bR ) & = -\left[ (\delta_{ij}-{\hat{R}}_i {\hat{R}}_j)\frac {k^2\cos (kR)}R - (\delta_{ij}-3{\hat{R}}_i {\hat{R}}_j)
\left(\frac {k\sin (kR)}{R^2} +\frac {\cos (kR)}{R^3} \right) \right]
\\
& = -F_{ij}^R \frac {\cos (kR)}{R} ,
\end{split}
\end{equation}
where $F_{ij}^R$ is the differential operator acting on the variable $R$ defined in (\ref{differential operator}).  It should be noted that the vacuum-induced correlation between the dipole moments of $B$ and $C$ appears in (\ref{interaction correlated dipoles}).Thus, using (\ref{induced dipole}) in (\ref{interaction correlated dipoles}), the electric-electric interaction energy becomes

\begin{equation}\label{ee interaction energy}
\Delta E^{ee}_{BC}= \sum_{\bk \lambda} \alpha_B^e(\omega) \alpha_C^e(\omega)  \langle 0 \rvert E_i(\bk \lambda ;\br_B ,t) E_j(\bkp \lambda '; \br_C ,t)\lvert 0 \rangle V_{ij}^{ee}(k , \br_C-\br_B ) ,
\end{equation}
showing a sharp relation between dispersion interactions and vacuum field correlations, as already mentioned \cite{Power-Thirunamachandran93b,Passante-Persico03}.

Explicit evaluation of (\ref{ee interaction energy}) yields the standard van der Waals interaction energy between two neutral ground-state atoms, as obtained by straightforward fourth-order perturbation theory \cite{Casimir-Polder48,Craig-Thirunamachandran98,Compagno-Passante05,Salam08,Salam10,Buhmann-II13}.
As it is well known, this dispersion energy scales with the distance as $R^{-6}$ in the near zone (London-van der Waals nonretarded regime) and as $R^{-7}$ in the far zone (Casimir-Polder retarded regime). The transition between the near and the far zone occurs at a distance around the main transition wavelength from the ground state of the atoms \cite{Casimir-Polder48,Craig-Thirunamachandran98,Compagno-Passante05}.
We wish to stress that the present evaluation of the dispersion energy in terms of vacuum spatial field correlation has some conceptual relevance, since it allows an insight of the origin of such interactions, showing that they can be interpreted partially from classical arguments, because quantum properties enter only in the existence of spatially-correlated vacuum fluctuations. Furthermore, in the far zone, where the atomic static (frequency independent) polarizability is involved, the physical interpretation is very clear: the spatial field correlation and the consequent dipoles correlation scales as $R^{-4}$, as (\ref{em EE correlation function}) and (\ref{induced dipole}) show, and the potential tensor scales as $R^{-3}$, yielding a far-zone interaction energy decreasing with the distance as $R^{-7}$.

Similar results can be analogously obtained for the magnetic-magnetic and electric-magnetic dispersion interactions between ground-state atoms, using the spatial correlation functions (\ref{em MM  F correlation function}) and (\ref{em EB  F correlation function},\ref{em BE  F correlation function}), respectively, together with the appropriate potential tensors $V_{ij}^{mm}(k , \bR )$ and $V_{ij}^{em}(k , \bR )$, given by \cite{Spagnolo-Dalvit07}

\begin{equation}\label{potential tensor mm}
V_{ij}^{mm}(k , \bR ) = -\left[ (\delta_{ij}-{\hat{R}}_i {\hat{R}}_j)\frac {k^2\cos (kR)}R - (\delta_{ij}-3{\hat{R}}_i {\hat{R}}_j)
\left(\frac {k\sin (kR)}{R^2} +\frac {\cos (kR)}{R^3} \right) \right] ,
\end{equation}

\begin{equation}\label{potential tensor em}
V_{ij}^{em}(k, \bR ) = \epsilon_{ij\ell} (\hat{R})_\ell \left[ \frac {k\sin (kR)}{R^2} - \frac {k^2 \sin (kr)}{R} \right] .
\end{equation}

An explicit evaluation yields, also for the magnetic-magnetic and electric-magnetic interactions \cite{Spagnolo-Dalvit07}, the known expressions as obtained by fourth-order perturbation theory \cite{Feinberg-Sucher70}. This shows that the relation between spatial field correlations and dispersion interactions holds also for magnetic-magnetic and electric-magnetic dispersion interactions.

\subsection{Three-body dispersion interactions}
\label{Three-body dispersion interactions}

The connection between spatial field correlations and dispersion interactions discussed in the previous subsection, can be extended to the non-additive three-body van der Waals and Casimir-Polder forces in terms of the dressed spatial correlations introduced in section \ref{sec:2}.

We consider three atoms, A, B and C, respectively located at $\br_A$, $\br_B$ and $\br_C$. The main physical point is that the presence of one atom, let say atom A, modifies the spatial field correlation evaluated at the points where the two other atoms B and C are located, as we have shown in section \ref{sec:2} for both a ground-state atom and an excited-state one. This change affects the interaction between atoms B and C, that now depends also on the presence of atom A: the spatial correlation function dressed by atom A, such as (\ref{dressed static correlation ground}) or (\ref{dressed static correlation excited}), is in this case involved in the expression analogous to (\ref{interaction correlated dipoles}). From now onwards we will consider only electric interactions between the atoms, so, in order to simplify the notation, we omit the apex $e$.

In the specific case of three ground-state atoms, the (electric-electric) interaction energy between B and C, in the presence of A, in analogy with (\ref{ee interaction energy}), is obtained by replacing the bare spatial correlation of the electric field with the dressed one given by (\ref{dressed static correlation ground}). Thus, it can be written down as

\begin{equation}
\label{interaction ground BC}
\Delta E_{BC} = \sum_{\bk \lambda \bkp \lambda '} \alpha_B(\wk )\alpha_C(\wkp ) \langle \tilde{g}_A \rvert E_i(\bk \lambda ;\br_B ) E_j(\bkp \lambda '; \br_C)\lvert \tilde{g}_A \rangle
V_{ij}(k, k'; \br_C-\br_B),
\end{equation}
where

\begin{equation}
\label{potential tensor symm}
V_{ij}(k,k'; \br_C-\br_B)= \frac 12 \left( V_{ij}(k, \br_C-\br_B)+V_{ij}(k', \br_C-\br_B) \right))
\end{equation}
is the symmetrized generalization of the potential tensor (\ref{potential tensor ee}), necessary since the dressed field correlation depends from the two wavenumbers $k$ and $k'$ (due to the presence of atom A, different field modes are now correlated).

Explicit evaluation of (\ref{interaction ground BC}) using (\ref{dressed static correlation ground}), yields two terms: a two-body contribution consisting on the direct interaction between atoms $B$ and $C$ (i.e. the same discussed in the previous subsection) plus a three-body term $\Delta_{BC}^3$ containing the effect of the presence of atom A on the dispersion interaction between B and C, stemming from the dressing term of
$\langle \tilde{g}_A \rvert E_i(\bk \lambda ;\br ,t) E_j(\bkp \lambda '; \brp ,t)\lvert \tilde{g}_A \rangle$ of the spatial correlation, that is the second term of (\ref{dressed static correlation ground}) \cite{Cirone-Passante97}.
The complete three-body dispersion potential is then obtained after a symmetrization over the role of the three atoms; after some lengthy algebraic calculations and introducing the imaginary frequency $u=-i\wk$, it can be cast in the following form

\begin{equation}
\label{three body ground state}
\begin{split}
\delta E_{ABC} & = \frac 23 \left( \Delta_{AB}^3 + \Delta_{BC}^3 + \Delta_{AC}^3\right)
\\
& = -\frac {\hbar c}{\pi}F_{ij}^\alpha F_{j\ell}^\beta F_{i\ell}^\gamma \frac 1{\alpha \beta \gamma}\int_0^\infty \! du \alpha_A(iu) \alpha_B(iu) \alpha_C(iu) e^{-u(\alpha +\beta +\gamma )} ,
\end{split}
\end{equation}
where $\alpha = \lvert \br_C -\br_B \rvert$, $\beta = \lvert \br_C -\br_A \rvert$ and $\gamma = \lvert \br_B -\br_A \rvert$ are the distances between the atoms \cite{Cirone-Passante97}. Eq. (\ref{three body ground state}) coincides with the known expression of the three-body component of the dispersion interaction as obtained by a standard sixth-order perturbative approach \cite{Aub-Zienau60,Salam10,Salam14,Salam16}, or also by field energy density considerations \cite{Power-Thirunamachandran85,Power-Thirunamachandran92} or using effective Hamiltonians \cite{Passante-Power98,Passante18}. Our approach, however, gives a clear physical insight on the origin of the nonadditive nature of such interactions, stressing the fundamental role played by dressed spatial correlations of vacuum fluctuations. Moreover, as we will discuss in the next section, our approach can be generalized to nonstationary conditions too.

In the far zone (retarded regime), that is when $\alpha , \beta , \gamma \gg \lambda_c$, with $\lambda_c$ a characteristic transition wavelength of the three atoms, only the static polarizabilities of the three atoms are involved in the three-body dispersion interaction, and, for a equilateral triangular configuration of the atoms with side $R$, it scales as $R^{-10}$ \cite{Aub-Zienau60,Power-Thirunamachandran85,Salam10}. This is fully consistent with our considerations after Eq. (\ref{correlation electric field ground 3}) that at a large distance from atom A the dressing part of the correlation function scales as $R^{-7}$, correctly yielding a potential energy scaling as $\sim R^{-7}R^{-3} \sim R^{-10}$ (the $R^{-3}$ term is from the potential tensor (\ref{potential tensor ee})).

\section{Dressed dynamical field correlations and dynamical three-body Casimir-Polder forces}
\label{sec:4}

The considerations on dressed field correlations in the previous sections refer to static (time-independent) situations. Also the case with one excited atom was considered for short times, so that its spontaneous decay can be neglected. In this section we will address dynamical nonstationary situations, when the system starts from a nonequilibrium configuration. We will address relevant aspects of the building up in time of nonlocal spatial field correlations, as well as dynamical time-dependent three-body Casimir-Polder interactions. We will show that relevant nonlocal features appear in the time evolution and build-up of the field spatial correlations, and how they manifest in the consequent time-dependent dispersion (van der Waals and Casimir-Polder) interactions  between three (or more) atoms. This suggests that dynamical dispersion interactions could provide a way to probe these nonlocal nonequilibrium field correlations, similarly to the stationary case of previous section.

We first concentrate on the nonstationary spatial field correlations when an atom (A) is present, in both cases of an initially bare ground-state atom and an initially bare excited-state atom. In such cases the initial state at $t=0$ is a nonequilibrium state, being an eigenstate of the unperturbed Hamiltonian and not of the total Hamiltonian: thus, they will evolve in time. The atom-field Hamiltonian is given by (\ref{Hamiltonian one atom}).

\subsection{Dynamical ground-state correlations}

We consider an atom A interacting with the quantum electromagnetic field via the dipolar interaction Hamiltonian (\ref{Hamiltonian one atom}). We assume that at $t=0$ the system is prepared in the noninteracting (bare) ground state $\lvert g_A, 0 \rangle$, where $\vert g_A\rangle$  is the ground state of the atom and $\lvert 0\rangle$ is the photon vacuum. Differently from section \ref{sec:2}, here we use a time-dependent approach, in order to study the time evolution of the correlation function starting from our nonequilibrium initial state. We work in the Heisenberg representation, and solve by iteration up to the second order the Heisenberg equations for the field operators

\begin{equation}\label{iterative solution}
\akl (t)  = \akl^{(0)}(t) + \akl^{(1)}(t) + \akl^{(2)}(t) + \ldots ,
\end{equation}
where the apex indicates the perturbative order. From the explicit iterative solution (\ref{iterative solution}) we can obtain the corresponding iterative solution for the electric field operator in the Heisenberg representation
\cite{Passante-Persico07,Power-Thirunamachandran99}

\begin{equation}\label{iterative solution 1}
\bE (\br ,t)  = \bE^{(0)}(\br ,t) + \bE^{(1)}(\br ,t) + \bE^{(2)}( \br ,t) + \ldots ,
\end{equation}
that we will exploit to obtain the equal-time electric-field spatial correlations in a two-level approximation for the atom, that is
$\bmu_A = \bmu^{eg}_AS_+ +\bmu^{ge}_AS_- $. Here, $S_+$, $S_-$ and $S_z$ are the pseudospin operators for atom A. Also, $\bmu^{eg}_A = \langle e_A \rvert \bmu_A \lvert g_A \rangle = {\bmu^{ge}_A}^*$
are the matrix elements of the atomic electric dipole operator between the excited state $\lvert e_A \rangle$ and the ground state $\lvert g_A \rangle$, with energy $E_e$ and $E_g$, respectively. We will however assume, as before, that the dipole matrix elements are real, so that $\bmu^{eg}_A = \bmu^{ge}_A$,  In the pseudospin formalism, the atomic Hamiltonian is $H_A=\hbar \wA S_z$, where $\wA = (E_e -E_g)/\hbar$.

Writing down the Heisenberg equations of motion for the field and atomic operators and solving them up to the second order in the atom-field coupling, we can finally obtain an explicit expression of the two-mode correlation function of the electric field operator in the initial bare ground state $\lvert g_A, 0 \rangle$ \cite{Passante-Persico07}

\begin{equation}
\label{correlation electric field ground dynamical 1}
\begin{split}
&\ C_{ij}^g(\bk \lambda , \bkp \lambda '; \br , \brp ,t) = \langle g_A,0 \rvert E_i(\bk \lambda ; \br ,t ) E_j(\bkp \lambda ' ; \brp ,t ) \lvert g_A,0 \rangle
\\
&= \frac {2\pi \hbar}{V}(\wk \wkp )^{1/2} (\ekl )_i (\eklp )_j  e^{i\bk \cdot (\br -\brp )} \delta_{\bk \bkp} \delta _{\lambda \lambda '}
\\
&- \frac {(2\pi )^2}{V^2} (\bmu_A^{eg} \cdot \ekl ) (\bmu_A^{eg} \cdot \eklp ) (\ekl )_i (\eklp )_j \wk \wkp 2\Re \Big\{
\frac {e^{i (\bk \cdot (\br -\br_A)-\wk t)}}{i(\wA +\wk )}
\\
&\times \left[ \big( F(\wk +\wkp ,t) - F^*(\wA -\wkp ,t)  \big) e^{i ( \bkp \cdot (\brp -\br_A)-\wkp t)}
-  \big( F(\wk -\wkp ,t) - F^*(\wA +\wkp ,t)  \big) \right.
\\
&\times \left. e^{-i ( \bkp \cdot (\brp -\br_A)-\wkp t)}  \right] + \text{c.c.}(\bk \rightleftharpoons \bkp , \br \rightleftharpoons \brp )
\Big\} ,
\end{split}
\end{equation}
where $\Re$ stands for the real part and we have defined the function

\begin{equation}
\label{F funcion}
F(x,t) = \int_0^t \! dt' e^{ixt'} = \frac {e^{ixt}-1}{ix} .
\end{equation}

Eq. (\ref{correlation electric field ground dynamical 1}) clearly shows that different field modes, initially uncorrelated, acquire with time a correlation consequent to their mutual interaction with the atom (contrarily to the case of the field bare vacuum).

We also obtain the two-point and equal-time nonstationary spatial correlation function for the complete electric field, given by

\begin{equation}
\label{correlation electric field ground dynamical 2}
\begin{split}
C_{ij}^g (\br ,\brp ,t) &= \langle g_A,0 \rvert E_i (\br ,t ) E_j (\brp ,t ) \lvert g_A,0 \rangle = \sum_{\bk \lambda \bkp \lambda '}  C_{ij}^g(\bk \lambda , \bkp \lambda '; \br , \brp ,t)
\\
&= \langle g_A,0 \rvert E_i^{(0)}(\br ,t) E_j^{(0)}(\brp ,t) \lvert g_A,0 \rangle + \langle g_A,0 \rvert E_i^{(1)}(\br ,t) E_j^{(1)}(\brp ,t) \lvert g_A,0 \rangle +
\\
&\ \ + \langle g_A,0 \rvert E_i^{(0)}(\br ,t) E_j^{(2)}(\brp ,t) \lvert g_A,0 \rangle + \langle g_A,0 \rvert E_i^{(2)}(\br ,t) E_j^{(0)}(\brp ,t) \lvert g_A,0 \rangle ,
\end{split}
\end{equation}
(first-order terms vanish) \cite{Power-Thirunamachandran93a,Power-Thirunamachandran99}. After some lengthy algebraic calculations, we find the following expressions of the single terms in (\ref{correlation electric field ground dynamical 2})

\begin{eqnarray}
\label{correlation electric field ground dynamical 3}
&\ &\langle g_A,0 \rvert E_i^{(0)}(\br ,t) E_j^{(0)}(\brp ,t) \lvert g_A,0 \rangle = \frac {2\pi \hbar}{V} \sum_{\bk j} (\ekl )_i (\eklp )_j \wk e^{i\bk \cdot (\br -\brp )},
\nonumber \\
&\ & \langle g_A,0 \rvert E_i^{(1)}(\br ,t) E_j^{(1)}(\brp ,t) \lvert g_A,0 \rangle
\nonumber \\
&\ & = (\mu_A^{eg})_m(\mu_A^{eg})_n F_{im}^R \left( \frac {e^{i\wA R/c}}{R}\right) F_{jn}^{R'} \left( \frac {e^{-i\wA R'/c}}{R} \right)
\theta (ct -R) \theta (ct -R') ,
\nonumber \\
&\ & \langle g_A,0 \rvert E_i^{(0)}(\br ,t) E_j^{(2)}(\brp ,t) \lvert g_A,0 \rangle + \langle g_A,0 \rvert E_i^{(2)}(\br ,t) E_j^{(0)}(\brp ,t) \lvert g_A,0 \rangle
\nonumber \\
&\ & = \frac {2\pi}{V} \Big\{ (\mu_A^{eg})_\ell (\mu_A^{eg})_m \sum_{\bk \lambda}(\ekl )_i (\ekl )_\ell  e^{i \bk \cdot \bR )} \frac {\wk}{\wk +\wA} F_{jm}^R \frac {1}{R} \left( e^{-i\wk R/c} -e^{-i(\wk +\wA )t}
e^{i\wA R/c} \right)
\nonumber \\
&\ & \times \theta (ct- R)
+\text{c.c.}( \bR \rightleftharpoons \bRp, A \rightleftharpoons B, i \rightleftharpoons j ) \Big\} ,
\end{eqnarray}
where $\bR = \br -\br_A$, $\bRp = \brp -\br_A$, and $\theta (x)$ is the Heaviside step function \cite{Passante-Persico07}. Also, $F_{ij}^R$ is the differential operator defined in (\ref{differential operator}).

The first term in (\ref{correlation electric field ground dynamical 2}), explicitly given in the first line of (\ref{correlation electric field ground dynamical 3}), coincides with the time-independent correlation in the bare ground state, as discussed in section \ref{sec:2} (see Eq. (\ref{em EE correlation function})); all other terms in (\ref{correlation electric field ground dynamical 2}) and (\ref{correlation electric field ground dynamical 3}) describe the time evolution of the correlation function during the dynamical self-dressing process of the atom, and are a main object of investigation in this section.

Some intriguing physical considerations are in order to physically understand the behavior of the dynamical spatial correlations we have obtained. For the sake of clarity, we separately consider the various contributions in the expressions above.

The second contribution in (\ref{correlation electric field ground dynamical 2}) and (\ref{correlation electric field ground dynamical 3}) is the product of the retarded dipole fields from atom A in points $\br$ and $\brp$ (the operator $\bE^{(1)}$ in (\ref{iterative solution 1})) \cite{Power-Thirunamachandran99}; the presence of the two $\theta$ functions expresses that this contribution to spatial correlation vanishes outside the light cone centered on atom A: it is different from zero only if both points $\br$ and $\brp$ are inside the causality sphere of A, in full agreement with relativistic causality. However, since $\lvert \br -\brp \rvert$ can be larger than $ct$ even if both $R = \lvert \br -\br_A \rvert$ and $R' = \lvert \brp -\br_A \rvert$ are smaller than $ct$, the correlation can be nonvanishing even for two points $r$ and $r'$ that are not not causally connected each other; this peculiar behavior is possible because the fields in both points are causally connected with the ``source'' atom A. This behavior clearly shows a nonlocal feature of dynamical spatial field correlations during the atomic dressing.

The contribution of the other terms in (\ref{correlation electric field ground dynamical 2}) and (\ref{correlation electric field ground dynamical 3}), involving the zeroth- and second-order electric field operators, shows a quite different and peculiar behavior. It is different from zero even if just one of the two points $\br$ and $\brp$ is inside the causality sphere of atom A, notwithstanding the other point can be outside of it. Mathematically, this is related to the fact that this term arises from a product of the second-order field generated by atom A (the operator $\bE^{(2)}$ in (\ref{iterative solution 1})) and the free field (the operator $\bE^{(0)}$ in (\ref{iterative solution 1})): the first one is causally connected with A while the second one is source independent (i.e., it is the free field at time $t$). In other words, this contribution to the spatial correlation function at a generic time $t$, can be nonvanishing even for points separated by a space-like interval, provided at least one of them is inside the causality sphere of A.

We can conclude that nonstationary field correlations, during the evolution of the atom-field system starting from our nonequilibrium configuration at $t=0$, have peculiar nonlocal features. An important point, in our opinion, is understanding whether such nonlocal features can manifest or be observed, at least indirectly, in observable physical quantities. In the next section we will show how these nonlocal features can indeed manifest in the dynamical three-body Casimir-Polder dispersion interactions between atoms or, in general, polarizable bodies.

\subsection{Dynamical excited-state correlations}

We now investigate the case when atom A, approximated as a two-level system, is initially in its bare excited state $\lvert e_A \rangle$, with field in its bare vacuum state $\lvert 0 \rangle$. Also in this case the initial state is a nonequilibrium state, and we expect a dynamical change of the spatial correlation function of the electric field; moreover, new aspects should appear with respect to the ground-state case, due to resonance effects related to the possibility of emission of a real photon.
The calculation proceeds similarly to the ground-state case of the previous subsection, using the same iterative-solution method for the Heisenberg equations of motion of the field and atom operators, and then evaluating the electric-field spatial correlation in the bare excited state of atom A and the field vacuum state, $\lvert e_A, 0 \rangle$. We consider time shorter than the lifetime of the excited state and thus we can neglect its spontaneous decay; however, contrarily to the ground-state case, a resonance pole in the frequency integrations is present in this case, yielding a new contribution to the correlation function.

After lengthy algebraic calculations, we obtain that the spatial correlation function of the electric field can indeed be separated in two contributions, a {\it resonant} one and a {\it nonresonant} one,

\begin{equation}
\label{correlation electric field ground dynamical 4}
\begin{split}
C_{ij}^e (\br ,\brp ,t) &= \langle e_A,0 \rvert E_i (\br ,t ) E_j (\brp ,t ) \lvert e_A,0 \rangle
\\
&= \langle E_i (\br ,t ) E_j (\brp ,t ) \rangle_{nr} + \langle E_i (\br ,t ) E_j (\brp ,t ) \rangle_{res} .
\end{split}
\end{equation}

The nonresonant term is found to be the same of that for the ground-state atom, Eqs. (\ref{correlation electric field ground dynamical 2}) and (\ref{correlation electric field ground dynamical 3}), with an opposite sign, that is $\langle E_i (\br ,t ) E_j (\brp ,t ) \rangle_{nr} = -C_{ij}^g (\br ,\brp ,t)$. The resonant term, resulting from the pole at $\wk = \wA$, is given by \cite{Passante-Persico07}

\begin{equation}
\label{correlation electric field ground dynamical 5}
\langle E_i (\br ,t ) E_j (\brp ,t ) \rangle_{res} = 2 (\mu_A^{eg})_m (\mu_A^{eg})_n F^{R}_{in} F^{R'}_{jm} \frac {\cos [(\wA (R - R')/c]}{RR'}\theta (ct-R) \theta (ct-R') .
\end{equation}
For the nonresonant contribution, the same physical considerations about its nonlocal features given in the previous subsection for the ground-state case apply, in particular that it is in general different from zero if one of the two points is inside the causality sphere of A, even if the other point is outside the causality sphere. In other words, the presence of the excited atom manifests itself in the two-point spatial correlation even if one point is outside the causality sphere of A, provided the other point is inside, and whatever the distance between the two points is. The resonant contribution
(\ref{correlation electric field ground dynamical 5}) shows, as expected, oscillations in space with a scale given by the atomic transition wavelength $c/\wA$; it is not vanishing provided both points are causally connected with the atom A, that is if $R,R' < ct$, even in the case they are not causally connected each other (that is, $\lvert \brp -\br \rvert > ct$).

We wish to stress that the nonlocal features of the dynamical spatial correlations we have discussed in this section are totally consistent with relativistic causality, of course, because correlations cannot transport information.

\subsection{Dynamical three-body Casimir-Polder interactions}

In this subsection we briefly analyze some consequence of the nonlocal character of the dynamical field correlations investigated in the previous subsections, specifically in dynamical Casimir-Polder three-body interactions, extending the stationary case of the previous section to nonstationary situations.

Our starting point is the relation between the spatial correlations of a quantum field, specifically of the electric field, and the dispersive two- and three-body Casimir-Polder forces, as discussed in section \ref{sec:3}. Here we mainly consider three-body interactions in nonstationary conditions, both in the case of three bare ground-state atoms and in the case of one bare excited atom and two ground-state atoms; we thus extend our methods and results for stationary three-body interactions, presented in section \ref{sec:3}, to the dynamical case, on the basis of the dynamical dressed field correlations obtained in the previous subsections.

We assume to have one atom (A) in $\br_A$ and two other ground-state atoms (B and C) in $\br_B$ and $\br_C$. We consider both cases of atom A initially in its bare ground state and in its bare excited state.

We first consider the case of atom A initially $(t=0)$ in its bare ground state and the field in the vacuum state. As already mentioned, this is a nonstationary condition, because bare states are not eigenstates of the complete Hamiltonian. We start evaluating the interaction energy between atoms B and C during the self-dressing of atom A assuming, as mentioned, the state of the interacting system field-(atom A) in the nonequilibrium bare state $\lvert g_A, 0 \rangle$. We evaluate this interaction energy exploiting the expression (\ref{interaction ground BC}), extended to the present nonstationary situation by substituting the two-mode correlation function of the electric field with its time-dependent expression, Eq. (\ref{correlation electric field ground dynamical 1}). After lengthy algebraic calculations, we can obtain the explicit analytical expression of the dynamical interaction energy, that at a generic time $t>0$ is given by \cite{Passante-Persico06}

\begin{equation}
\label{three body ground state dynamical 1}
\begin{split}
&\ \Delta E_A(B,C;t) = -\frac 1{2\pi} (\mu_A^{eg})_n (\mu_A^{eg})_p F^{\alpha}_{\ell m} F^{\beta}_{mp} F^{\gamma}_{\ell n} \frac 1{\alpha \beta \gamma}
\Re\Bigg\{ \int_0^\infty \! \! d\omega \frac{\alpha_B(\omega )}{\omega +\wA}
\\
&\times \Big[
2\big( \alpha_C(\omega ) e^{-i\omega \beta /c} - \alpha_C(\wA  ) e^{-i\wA \beta /c} e^{-i(\wA +\omega )t} \big)
\sin \left( \frac {\omega \gamma}{c} \right) \cos \left( \frac {\omega \alpha}{c} \right))  \theta (ct -\beta )
\\
&\ + \sin \left( \frac {\omega \gamma}{c} \right) \big( \alpha_C (\omega ) e^{-i\omega (\alpha +\beta )/c} -\alpha_C (\wA ) e^{-i\wA (\alpha +\beta  )/c} e^{-i(\wA +\omega )t} \big)  \big) \theta (ct - (\alpha +\beta ))
\\
&\ + \sin \left( \frac {\omega \gamma}{c} \right) \big( \alpha_C (\omega ) e^{-i\omega \lvert \beta -\alpha \rvert /c} -\alpha_C (\wA ) e^{-i\wA \lvert \beta -\alpha  \rvert /c} e^{-i(\wA +\omega )t} \big)
\sgn (\beta - \alpha ) \theta (ct - \lvert \beta -\alpha \rvert ) \Big]
\\
&\ +\text{c.c.} (B \rightleftharpoons C, \beta \rightleftharpoons \gamma ) \Bigg\} ,
\end{split}
\end{equation}
where $\sgn (x)$ is the function giving the sign of the variable $x$, $\theta (x)$ is the Heaviside step function and $\alpha$, $\beta$, $\gamma$ are the distances between the atoms defined in subsection \ref{Three-body dispersion interactions}.

Several physical considerations relevant to nonlocality and causality can be done starting from (\ref{three body ground state dynamical 1}), in different specific situations.

Let us first assume that both atoms B and C are inside the causality sphere of A. In such a case it is possible to show from (\ref{three body ground state dynamical 1}) that, after a transient period when the interaction energy is time-dependent, it settles to a stationary value \cite{Passante-Persico06}, that coincides with that found with the time-independent approach of Section \ref{Three-body dispersion interactions}. This also shows the self-consistency of our time-dependent approach.

A remarkable and unexpected result is when at least one of the two atoms B and C is outside the light-cone of A, i.e. when $\beta > ct$ and/or $\gamma > ct$; evaluating (\ref{three body ground state dynamical 1}) in this case it follows that $\Delta E_A(B,C;t)$ can be nonvanishing, and that the interaction between B and C can be affected by A, clearly showing a nonlocal behavior of the three-body dispersion energy due to the nonlocality of the dynamical correlation function of the electric field; more details on the ranges of the relevant parameters $\alpha$, $\beta$ $\gamma$, $t$ in which this occurs can be found in \cite{Passante-Persico06}.

A relevant quantity worth of investigation is the symmetrized dynamical interaction energy, given by

\begin{equation}
\label{three body ground state dynamical symmetrized}
\Delta(A,B,C,t) = \frac 23 \left( \Delta E_A(B,C;t) + \Delta E_B(A,C;t) + \Delta E_C(A,B;t) \right) ,
\end{equation}
where the role of the three atoms is exchanged in the symmetrization (for a more detailed discussion of the difference between the time-dependent interaction energies (\ref{three body ground state dynamical 1}) and (\ref{three body ground state dynamical symmetrized}), in particular in relation with their measurement, see \cite{Rizzuto-Passante07}). Its expression can be directly obtained from (\ref{three body ground state dynamical 1}). We will discuss here only its main features relevant to the indirect manifestation of the nonlocality of dressed field correlations in three-body dispersion interactions, aspect we are mainly interested to.

Firstly, after a transient, it is possible to show that the dynamical interaction energy (\ref{three body ground state dynamical symmetrized}) settles to its stationary equilibrium value (similarly to (\ref{three body ground state dynamical 1})), as given by (\ref{three body ground state}) or by a direct sixth-order perturbative calculation \cite{Aub-Zienau60,Power-Thirunamachandran85,Salam10,Salam14}.

Secondly, although (\ref{three body ground state dynamical symmetrized}) vanishes if each atom is outside the causality sphere of the other two, i.e. $\alpha, \beta, \gamma >ct$, it nevertheless shows relevant nonlocal aspects: for example, for times such that one atom (A, for example) is not causally connected with the other two atoms (B and C), while B and C are causally connected to each other, situation represented by the conditions $\beta > ct$, $\gamma > ct$, $\alpha < ct$, the dynamical three-body interaction (\ref{three body ground state dynamical 1}) is not vanishing. In other word, this observable interaction energy manifests a nonlocal behaviour, ultimately related to the nonlocal features of the dynamical spatial correlations of the electric field, since it is not vanishing even if one atom is outside the causality sphere of the other two \cite{Passante-Persico06,Rizzuto-Passante07}.

Similar considerations can be done in the nonstationary case in which one of the three atoms (A) is initially in its bare excited state with the field in its vacuum state, using the dynamical excited state field correlations given in (\ref{correlation electric field ground dynamical 4}). In this case, the main difference with the previous case is the presence of the resonant term
(\ref{correlation electric field ground dynamical 5}) in the spatial correlation function of the electric field.
This extra term finally yields the following additional contribution to the dynamical three-body interaction energy \cite{Passante-Persico07}

\begin{equation}
\label{three body excited state resonant}
\begin{split}
&\Delta E_{res}(A,B,C,t) = - (\mu_A^{ge})_n (\mu_A^{ge})_p \alpha_B (\w0 ) \alpha_C (\w0 )2\Re \left( F_{\ell n}^\gamma F_{mp}^\beta
\frac {e^{-i\w0 (\beta +ct)/c}}{\beta \gamma}  \right)
\\
&\times F_{\ell m}^\alpha \left( \frac {\cos (\w0 \alpha /c)}{\alpha} \right) \theta (ct-\beta ) \theta (ct- \gamma ),
\end{split}
\end{equation}
where the same notation of the previous cases have been used. The resonant contribution (\ref{three body excited state resonant}) at time $t$ is not vanishing only if both atoms B and C are inside the light cone of A, i.e. $\beta < ct$ and $\gamma < ct$ even if they can be causally disconnected each other, for example if $\lvert \br_B - \br_C \rvert > ct$, showing also in this case some nonlocal aspect of the resonant term of the interaction energy.

Finally, we wish to mention that similar considerations about the nonlocal features of the three-body component of the dynamical dispersion interaction between three atoms can be obtained by evaluating them exploting effective Hamiltonians \cite{Passante-Power98,Passante-Rizzuto21}, as discussed in detail in \cite{Rizzuto-Passante07,Passante-Persico06,Passante-Persico07} also with reference to the measurement of the three-body component of dispersion interactions which involves the overall system and thus it is intrinsically nonlocal.

\section{Conclusions}
\label{sec:5}
In this paper we have reviewed several aspects of spatial correlations and their nonlocal features of quantum fields, mainly the electromagnetic field in the Coulomb gauge, both in their bare vacuum state and in dressed states when a ground- or excited-state atom is present. We have also considered dynamical (time-dependent) conditions when the interacting atom-field system starts from a nonequilibrium state, such as the bare ground or excited states of the atom, with the field in its vacuum state. We have discussed in detail the nonlocality of the spatial correlations of the quantum electric field in stationary and nonstationary situations, and their relation to two- and three-body dispersion interactions between atoms (van der Waals and Casimir-Polder). Specifically, we have shown that such observable interatomic interactions allow, both in stationary and nonstationary configurations, to gain an indirect evidence on the existence of these nonlocal spatial field correlations, and their physical properties.

\vspace{6pt}

\section*{Acknowledgements}
R.P. and L.R. acknowledge financial support from the FFR2021 and FFR2023 grants from the University of Palermo, Italy.

\bibliography{biblio}

\end{document}